\documentclass{ws-ijmpa}
\usepackage[super,compress]{cite}
\usepackage{graphicx}
\usepackage{hyperref}
\hypersetup{colorlinks,urlcolor=black,citecolor=black,linkcolor=black,filecolor=black}
\usepackage{breakurl}
\usepackage{pict2e}

\newcommand{\be}{\begin{equation}}
\newcommand{\ee}{\end{equation}}
\newcommand{\bea}{\begin{eqnarray}}
\newcommand{\eea}{\end{eqnarray}}
\newcommand{\nn}{\nonumber}

\newcommand{\vB}{\mathbf{B}}
\newcommand{\vE}{\mathbf{E}}
\newcommand{\vH}{\mathbf{H}}
\newcommand{\vD}{\mathbf{D}}
\newcommand{\ve}{\mathbf{e}}
\newcommand{\vk}{\mathbf{k}}
\newcommand{\ct}{\cos\theta}

\begin{document}
\markboth{A. Rebhan, G. Turk}{Polarization effects in light-by-light scattering}

%
\catchline{}{}{}{}{}
%

\title{Polarization effects in light-by-light scattering:\\ Euler-Heisenberg versus Born-Infeld%
\footnote{
Preprint of an article published in \href{http://www.worldscientific.com/worldscinet/ijmpa}{International Journal of Modern Physics A} 32 (2017) 1750053, DOI: 10.1142/S0217751X17500531
\copyright\ 2017 World Scientific Publishing Company 
\url{http://www.worldscientific.com/worldscinet/ijmpa}
}
}

\author{Anton Rebhan and G\"unther Turk}

\address{Institut f\"ur Theoretische Physik, Technische Universit\"at Wien,\\ Wiedner Hauptstra\ss e 8-10, A-1040 Vienna, Austria\\
anton.rebhan@tuwien.ac.at}



\maketitle

\begin{history}
\received{30 January 2017}
\end{history}

\begin{abstract}
The angular dependence of the differential cross section of
unpolarized light-by-light scattering summed over final polarizations
is the same in any low-energy effective theory of quantum electrodynamics
and also in Born-Infeld electrodynamics.
In this paper we derive general expressions for polarization-dependent
low-energy scattering amplitudes, including a hypothetical parity-violating situation.
These are evaluated for quantum electrodynamics
with charged scalar or spinor particles, which give strikingly different 
polarization effects. Ordinary
quantum electrodynamics is found to exhibit rather intricate polarization patterns
for linear polarizations,
whereas supersymmetric quantum electrodynamics 
and Born-Infeld electrodynamics give particularly simple forms.
\end{abstract}



\section{Introduction}

In 1935, long before quantum electrodynamics (QED) was in place as the fundamental theory of
electromagnetic interactions, Euler and Kockel\cite{Euler:1935zz,Euler:1936} evaluated its implications
on the nonlinear phenomenon of the scattering of light by light at energies
below the electron-positron pair creating threshold.\footnote{The ultrarelativistic limit was
derived immediately thereafter, in 1936, by Akhiezer, Landau, and Pomeranchuk\cite{Akiezer1936,Akiezer1937};
the complete leading-order result was worked out finally by Karplus and Neuman\cite{Karplus:1950zz}.}
The underlying effective action quartic in the electromagnetic field strength tensor that Euler
and Kockel had obtained was then generalized
to all orders in the famous paper by Heisenberg and Euler\cite{Heisenberg:1935qt}, and extended
to the case of charged scalar particles by Weisskopf\cite{Weisskopf:1936}, all in 1936.\footnote{See
Ref.~\citen{Dittrich:2000zu,Dunne:2004nc} for a review of further developments.}

A rather different action for nonlinear electrodynamics was proposed in 1934 by Born and Infeld\cite{Born:1934gh},
whose aim was to eliminate the infinite self-energy of charged particles in classical electrodynamics
and which for some time also carried the hope of taming the infinities of quantum field theory.
Born-Infeld (BI) electrodynamics leads to light-by-light scattering already at the classical level,
which was studied by Schr\"odinger in the 1940's.\cite{Schrodinger:1942,Schrodinger:1943}
Remarkably, this theory surfaced again in string theory as the effective action of Abelian
vector fields in open bosonic strings \cite{Fradkin:1985qd}.\footnote{There are supersymmetric extensions of the BI Lagrangian which
differ in terms beyond quartic order in the field strength,\cite{Cecotti:1986gb} however
the full supersymmetry of ten-dimensional superstrings again
singles out the original form.\cite{Tseytlin:1986ti,Metsaev:1987qp,Tseytlin:1999dj}
}
In fact, many of its
curious properties can be understood from a string theoretic point of view\cite{Gibbons:1997xz,Gibbons:2000xe}.


In this paper we revisit the polarization effects in low-energy light-by-light scattering that have been
worked out previously in ordinary (spinor) QED\cite{Karplus:1950zz} for circular polarizations, and we
generalize to the most generic low-energy effective action quartic in field strengths, including
also a parity (and CP) violating term. 

With circular polarizations the various differential cross sections have a rather simple form,
where only the magnitude, but not the angular dependence, depends on
the parameters of the low-energy effective action, {\it i.e.},
on the matter content of the fundamental theory.
However, with linear polarizations one obtains also widely different
angular dependences. Moreover, P and CP odd effects are separated
from the other contributions when linearly polarized states are considered.

We admit that our study is mostly of mere academic interest. We are not aware of any
concrete theory in the current literature that would lead to the P and CP odd term in the effective action for low-energy
light-by-light scattering that we are considering.\footnote{One way to produce such a term would be a
coupling of photons to axions and dilatons in a CP-breaking background.}
However, light-by-light scattering is one of the current research topics 
in high intensity laser physics\cite{Marklund:2008gj} and polarization effects are of
great relevance there, see {\it e.g.} Refs.~\citen{Heinzl:2006xc,Karbstein:2015xra} 
where it has been proposed that the effect of vacuum birefringence\cite{Klein:1964,Baier:1967zzc,Brezin:1971nd}
may be tested in counter-propagating laser beams
(see also Ref.~\citen{Fouche:2016qqj} for more general tests of
nonlinear electrodynamics).

%

\section{Low-energy effective actions for light-by-light scattering}

In the limit of photon energies much smaller than the masses of charged particles, the latter
can be integrated out, 
yielding a gauge and Lorentz invariant effective action that is constructed from the field strength tensor
and where the leading terms involve the latter without further derivatives.
In an Abelian theory, it is well known that there are only two independent Lorentz (pseudo-)scalars, which we define as
\bea
\mathcal F&=&\frac14 F_{\mu\nu}F^{\mu\nu}\equiv-\frac14 \tilde F_{\mu\nu}\tilde F^{\mu\nu}
=-\frac12(\vE^2-\vB^2),\nonumber\\
\mathcal G&=&\frac14 F_{\mu\nu}\tilde F^{\mu\nu}=-\vE\cdot\vB,
\eea
with $\tilde F^{\mu\nu}=\frac12\varepsilon^{\mu\nu\rho\sigma}F_{\rho\sigma}$ in
the conventions of a mostly-minus metric and $\varepsilon^{0123}=+1$.
More complicated Lorentz scalars such as {\it e.g.} $F^{\mu\nu}F_{\nu\lambda}F^{\lambda\rho}F_{\rho\mu}$
can always be reduced to combinations of $\mathcal F$ and $\mathcal G$. (This is most easily
understood by the fact that rotational invariance already restricts to three possible invariants,
namely $\vE^2$, $\vB^2$, and $\vE\cdot\vB$. Boost invariance reduces these to two only.) 

The most general low-energy effective action for elastic light-by-light scattering
therefore has the form
\be\label{L4lowen}
\mathcal L^{(4)}_{\rm low\,en.}=c_1 \mathcal F^2+c_2 \mathcal G^2+c_3 \mathcal F \mathcal G. 
\ee
If one furthermore demands invariance under P and CP transformations, the third term
is forbidden. It is kept here for generality and in order to see what features in
the scattering cross section it would give rise to.


\begin{table}[t]
\tbl{Coefficients $c_{1,2}/C$ with $C=\alpha^2/m^4$. 
(In the BI case we have $c_1=c_2=1/(2b^2)$.)}
{\begin{tabular}{@{}ccc@{}} 
\toprule
 & $c_1/C$ & $c_2/C$ \\ 
\colrule
scalar QED& 7/90 & 1/90 \\
spinor QED& 8/45 & 14/45\\
supersymmetric QED & 1/3 & 1/3 \\
\botrule
\end{tabular} \label{tabc12}}
\end{table}

The one-loop contributions to $c_1$ and $c_2$ in spinor and scalar QED have been first obtained
by Euler and Kockel\cite{Euler:1935zz,Euler:1936} and Weisskopf\cite{Weisskopf:1936}, respectively,
and are reproduced in Table \ref{tabc12}.

The case of low-energy light-by-light scattering in supersymmetric QED was discussed in
Ref.~\citen{Duff:1979bk} as an illustration of a connection between self-duality,
helicity, and supersymmetry discovered initially in the context of supergravity.\cite{Grisaru:1976vm}
In supersymmetric QED the matter content is given by two charged scalar particles in addition
to the charged Dirac fermion. As shown in Table \ref{tabc12}, adding twice the contributions
of scalar QED to spinor QED leads to $c_1=c_2$.
This corresponds
to self-duality of the quartic term,\cite{Duff:1979bk} since then
one has
\be
\mathcal L^{(4)}_{\rm low\,en.,susy}\propto \mathcal{F}^2+\mathcal{G}^2=(\mathcal{F}+i\mathcal{G})(\mathcal{F}-i\mathcal{G})
\ee
with
\be
\mathcal{F}\pm i\mathcal{G}=\frac12 (F^\pm_{\mu\nu})^2,\quad F^\pm_{\mu\nu}:=\frac1{2}(F_{\mu\nu}\pm
i\tilde F_{\mu\nu}).
\ee

The same self-dual form at quartic order in the field strength is
found in
Born-Infeld electrodynamics, which is given by
\bea
\mathcal{L}^{\rm BI}&=&-b^2\sqrt{-\det\left(g_{\mu\nu}+\frac1b F_{\mu\nu}\right)}\nn\\
&=&-b^2 \left(1+2 b^{-2}\mathcal F-b^{-4}\mathcal G^2\right)^{1/2}=
-b^2-\mathcal F+\frac1{2b^2}(\mathcal{F}^2+\mathcal{G}^2)+O(b^{-4}),
\eea
where the parameter $b$ has the meaning of a limiting field strength (in static situations).
In fact, Born-Infeld electrodynamics 
features a nonlinear generalization of Hodge duality invariance that was pointed out already in 1935
by Schr\"odinger\cite{Schrodinger:1935oqa}, namely an invariance under
the transformations $(\vE+i\vH)\to e^{i\alpha}(\vE+i\vH)$,
$(\vD+i\vB)\to e^{i\alpha}(\vD+i\vB)$ with $\vD=\partial\mathcal{L}/\partial\vE$, $\vH=-\partial\mathcal{L}/\partial\vB$.
(See Refs.~\citen{BialynickiBirula:1984tx,Gibbons:1995cv,Gaillard:1997zr} for further discussions.)

Note, however, that supersymmetric Euler-Heisenberg Lagrangians\cite{Kuzenko:2007cg} are in general different from
Born-Infeld Lagrangians and their supersymmetric generalizations\cite{Cecotti:1986gb} beyond the quartic term
in the electromagnetic field strength.

\section{Scattering Amplitudes}

The amplitude for elastic photon scattering with given photon momenta 
$k_1^\mu,\ldots,k_4^\mu$ (with $\sum_i k_i=0$) and polarizations $\epsilon_1,\ldots,\epsilon_4$
is obtained from (\ref{L4lowen}) by
\be\label{M1234}
\mathcal{M}_{\epsilon_1\epsilon_2\epsilon_3\epsilon_4}(k_1,k_2,k_3,k_4)
=\left[\prod_{j=1}^4 i(k_j^\rho \epsilon_j^\sigma-k_j^\sigma \epsilon_j^\rho)
\frac{\partial}{\partial F_{\rho\sigma}}\right]
i\mathcal L^{(4)}.
\ee
This produces 24 terms for each of the terms in (\ref{L4lowen}).\footnote{As already noted in
Ref.~\citen{Davila:2013wba}, this immediately shows that the prescription given in the textbook by Itzykson and
Zuber\cite{IZ} has an error in the combinatorics. However, while the formula for $\mathcal M$
in Eq.~(7-97) of Ref.~\citen{IZ} misses a factor 24, the final result for $d\sigma/d\Omega$ given therein
is correct (but the resulting total cross section $\sigma$ contains a typo, see below for the correct value).}

For linear polarizations, we can write $\epsilon=(0,\ve)$ with a real unit vector $\ve$ orthogonal to $\vk$. 
We denote
$\ve_i$ and $\ve_o$ for the directions in and out of the plane of the scattering, respectively,
such that $\ve_i$, $\ve_o$ and $\vk/|\vk|$ form a right-handed orthogonal basis of unit vectors.
For circular polarizations, we introduce the complex unit vectors
\be\label{vepm}
\ve_{\pm}=\frac1{\sqrt2} (\ve_i \pm i\ve_o),
\ee
where the index $+/-$ denotes positive/negative helicities.\footnote{In optics,
positive helicity is often denoted as left-handed circular polarization, which is
at variance with particle physics as well as IEEE conventions.}
Note that $\ve_{\pm}$ are orthonormal in the sense $\ve_{\pm}^*\cdot\ve_{\pm}=\ve_{\mp}\cdot\ve_{\pm}=1$, $\ve_{\mp}^*\cdot\ve_{\pm}=\ve_{\pm}\cdot\ve_{\pm}=0$.

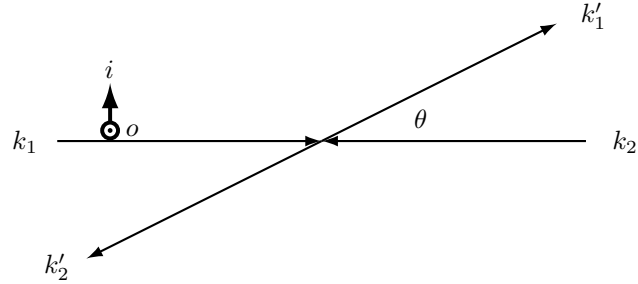
\begin{figure}[t]
\begin{picture}(400,100)
\thicklines
 \put(100,50){\vector(1,0){100}}
 \put(300,50){\vector(-1,0){100}}
 \put(200,50){\vector(2,1){89}}
 \put(200,50){\vector(-2,-1){89}}
 \put(83,47){$k_1$}
 \put(310,47){$k_2$}
 \put(95,0){$k'_2$}
 \put(298,95){$k'_1$}
 \put(235,55){$\theta$}
 \put(120,54){\circle{1}}
 \linethickness{1.5pt}
 \put(120,57){\vector(0,1){15}}
 \put(118,74){$i$}
 \put(120,54){\circle{6}}
 \put(126,52){$o$}
\end{picture}
\caption{Kinematics of photon-photon collisions in the center-of-mass system.\label{fig:kinematics}}
\end{figure}

The scattering amplitudes, being Lorentz scalars, can be expressed in terms of the Mandelstam variables $s,t,u$.
In the center-of-mass system, the only variables are $\omega=|\vk|$ and one polar angle
$\theta$ (see Fig.~\ref{fig:kinematics}), which are related to
the Mandelstam variables by
\bea
&&s=(k_1+k_2)^2=4\omega^2,\nn\\ 
&&t=(k_1-k'_1)^2=-2\omega^2(1-\ct)=-4\omega^2 \sin^2\textstyle{\frac{\theta}2},\nn\\
&&u=(k_1-k'_2)^3=-2\omega^2(1+\ct)=-4\omega^2 \cos^2\textstyle{\frac{\theta}2},
\eea
where $k'_1=-k_3$ and $k'_2=-k_4$. (Note that in the case of complex polarization
vectors the final polarizations in $\gamma\gamma\to
\gamma\gamma$ are given
by $\epsilon'_1=\epsilon_3^*$, $\epsilon'_2=\epsilon_4^*$.)

Evaluating (\ref{M1234}) for circular polarizations we obtain
\bea
-i\mathcal M_{++++}&=&\textstyle{\frac12} (c_1 - c_2 + i c_3)(s^2+t^2+u^2)\nn\\
&=&4  (c_1 - c_2 + i c_3) \omega^4 (3 +  \cos^2 \theta),\;\\
\mathcal M_{+++-}&=&\mathcal M_{++-+}=\mathcal M_{+-++}=\mathcal M_{-+++}=0,\\ 
-i\mathcal M_{++--}&=&\textstyle{\frac12}(c_1+c_2)s^2=8(c_1+c_2)\omega^4,
\label{M++--}\\
-i\mathcal M_{+-+-}&=&\textstyle{\frac12}(c_1+c_2)t^2=8(c_1+c_2)\omega^4\sin^4(\theta/2),\\
-i\mathcal M_{+--+}&=&\textstyle{\frac12}(c_1+c_2)u^2=8(c_1+c_2)\omega^4\cos^4(\theta/2),
\eea
and all other amplitudes are obtained by complex conjugation which flips all helicities, e.g.\
$\mathcal M_{----}=\mathcal M_{++++}^*$.

For the coefficients $c_i$ corresponding to spinor QED (see Table \ref{tabc12}), this reproduces
the low-energy result given in Refs.~\citen{Karplus:1950zz,Liang:2011sj} (as shown in the latter,
amplitudes with an odd number of $+$ or $-$ helicities start to contribute at order $\omega^6/m^6$).

Notice that the P and CP odd contribution proportional to $c_3$ shows up only in the amplitude
$M_{++++}=\mathcal M_{++++}^*$, corresponding to scattering with
polarizations $++\to--$ and $--\to++$, 
where it introduces a phase in the otherwise purely imaginary expression.

The amplitudes for the linear polarizations in and out of the collision plane read
\bea
-i\mathcal M_{iiii}&=&\textstyle{\frac12}c_1(s^2+t^2+u^2)=4c_1\omega^4 (3 +  \cos^2 \theta),\\
-i\mathcal M_{iiio}&=&-i\mathcal M_{iioi}=-i\mathcal M_{ioii}=-i\mathcal M_{oiii}=-\textstyle{\frac14}c_3(s^2+t^2+u^2) \nn\\
&=&-2c_3\omega^4 (3 +  \cos^2 \theta),\\
-i\mathcal M_{iioo}&=&-\textstyle{\frac12}c_1 s^2+\textstyle{\frac12}c_2(t^2+u^2)=
-8c_1\omega^4+4c_2\omega^4(1+  \cos^2 \theta),\\
-i\mathcal M_{ioio}&=&-\textstyle{\frac12}(c_1+c_2)su-\textstyle{\frac14}(c_1-c_2)(s^2+t^2+u^2) \nn\\
&=&[4(c_1+c_2)(1+\ct)+2(c_2-c_1 )(3+\cos^2 \theta)]\omega^4, \nn\\
&=&[11 c_2 -3 c_1 + 4 (c_1 + c_2) \ct + (c_2-c_1 ) \cos 2\theta]\omega^4 \\
-i\mathcal M_{iooi}&=&-\textstyle{\frac12}(c_1+c_2)st-\textstyle{\frac14}(c_1-c_2)(s^2+t^2+u^2) \nn\\
&=&[4(c_1+c_2)(1-\ct)+2(c_2-c_1 )(3+\cos^2 \theta)]\omega^4 \nn\\
&=&[11 c_2 -3 c_1 - 4 (c_1 + c_2) \ct + (c_2-c_1 ) \cos 2\theta]\omega^4.
\eea
All amplitudes are invariant under flipping all linear polarizations $i\leftrightarrow o$, which
fixes those not explicitly given.
(Note that also the amplitudes with linear polarizations can be expressed solely in terms of squares of
Mandelstam variables by rewriting $su=(t^2-s^2-u^2)/2$ and $st=(u^2-s^2-t^2)/2$.)

In contrast to the case of circular polarizations, all amplitudes for linear polarizations are
purely imaginary and the P and CP odd contribution is separated in the amplitudes with an odd number
of $i$ or $o$ polarizations.

In supersymmetric QED and in Born-Infeld electrodynamics, where $c_3=0$ and $c_1=c_2=1/(2b^2)$, 
the scattering amplitudes simplify in that $\mathcal M_{++++}^{\rm BI/susy}=\mathcal M_{----}^{\rm BI/susy}=0$, 
because $\mathcal{L}^{(4)}\propto (F^+_{\mu\nu})^2 (F^-_{\mu'\nu'})^2$ requires an equal
number of $+$ and $-$ helicities.
The amplitudes with mixed linear
polarizations also simplify and take the special forms
\bea
-i\mathcal M_{iioo}^{\rm BI/susy}&=&-2 b^{-2} \omega^4 \sin^2\theta,\nn\\
-i\mathcal M_{ioio}^{\rm BI/susy}&=&8 b^{-2} \omega^4 \cos^2(\theta/2),\nn\\
-i\mathcal M_{iooi}^{\rm BI/susy}&=&8 b^{-2}  \omega^4 \sin^2(\theta/2).
\eea

\section{Differential Cross Sections}

The final expression for the differential cross section reads
\be
\frac{d\sigma}{d\Omega}=\frac1{(16\pi)^2\omega^2}|\mathcal{M}_{\epsilon_1\epsilon_2\epsilon'^*_1\epsilon'^*_2}(k_1,k_2,-k'_1,-k'_2)|^2
\ee
in the center-of-mass system, to which we will stick in what follows.

Let us just point out that with the results given above in terms of Mandelstam variables,
an equivalent, frame-independent expression for light-by-light scattering is given by
\be
\frac{d\sigma}{dt}=\frac1{16\pi s^2}|\mathcal M|^2,
\ee
which would be useful for describing the scattering of photons with unequal energies.

\subsection{Unpolarized inital states with summation over final polarizations}

\begin{figure}[t]
\centerline{\includegraphics[width=.65\textwidth]{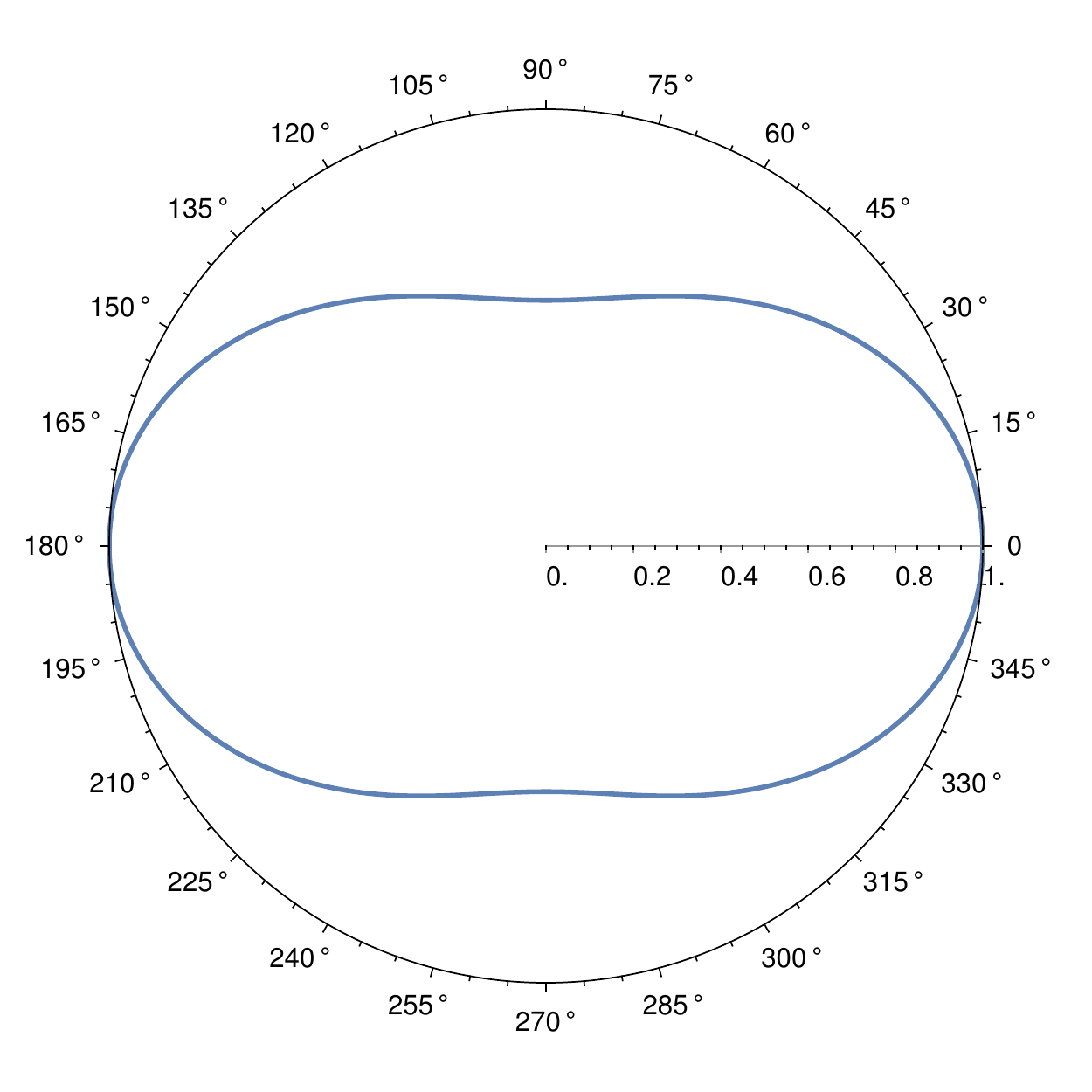}}
\caption{Polar plot of the universal form of $d\sigma/d\Omega\propto (3+\cos^2\theta)^2$ for unpolarized light-by-light scattering at leading order in $\alpha$ and $\omega/m$. 
The same angular dependence of $d\sigma/d\Omega$ appears for photon polarizations
$++\to--$ and $--\to++$
and also in all
parity-violating contributions involving $c_3$ such as $d\sigma/d\Omega$ for $ii\to io$.
\label{f1}}
\end{figure}

The unpolarized differential cross section for low-energy light-by-light scattering, averaged over inital polarizations and summed over final polarizations,
reads 
\be\label{sunpol}
\frac{d\sigma^{\rm unpol.}}{d\Omega}=\frac{\omega^6}{64\pi^2}\left(3c_1^2-2c_1c_2+3c_2^2+2c_3^2\right)(3 +  \cos^2 \theta)^2.
\ee
Evidently, this result has a universal dependence on the scattering angle, which is displayed as a polar plot
in Fig.~\ref{f1}.

In ordinary spinor QED (see Table \ref{tabc12}), this gives the well-known result\cite{Euler:1936,Karplus:1950zz,LL4,IZ}
\be
\frac{d\sigma^{\rm unpol.}_{\rm QED}}{d\Omega}=\frac{139 \alpha^4 \omega^6}{(180\pi)^2 m^8}(3 +  \cos^2 \theta)^2.
\ee
Replacing electrons by two charged scalar fields of the same mass as electrons would amount to replacing
the factor 139 by 34. Scalar QED, even with the same number of degrees of freedom as
ordinary QED, thus turns out to be much
less efficient in scattering light by light in the low-energy region.
Finally, supersymmetric QED would have a factor 225 in place of 139.

The total cross section is given by
\be
\sigma=\frac12 \int d\Omega \frac{d\sigma}{d\Omega},
\ee
where the factor $1/2$ is due to having identical particles in the final state.
(Alternatively, one could do without this symmetry factor and integrate over only one hemisphere\cite{IZ}.)
This yields
\be
\sigma(\gamma\gamma\to\gamma\gamma)^{\rm unpol.}=\frac{7(3c_1^2-2c_1c_2+3c_2^2+2c_3^2)\omega^6}{20\pi}.
\ee
In ordinary QED one obtains
\be
\sigma(\gamma\gamma\to\gamma\gamma)^{\rm unpol.}_{\rm QED}=\frac{973 \alpha^4 \omega^6}{10125 \pi m^8}
\ee
in agreement with Refs.~\citen{Euler:1936,Karplus:1950zz,LL4}.\footnote{Ref.~\citen{IZ} contains a typo here: 
the factor $\frac{56}{11}$
in Eq.~(7-101) should read $\frac{56}{5}$.}

\subsection{Final polarization with initial unpolarized photons}

When the polarizations of the photons after the scattering of initially unpolarized photons are measured,
the angular dependence of the differential cross section is in general different from (\ref{sunpol}).

Separating the contributions of equal and opposite circular polarizations in the final state, we obtain
\bea\label{sunpol++}
\frac{d\sigma^{{\rm unpol.\to ++}}}{d\Omega}=\frac{\omega^6}{2(16\pi)^2}
&&\Bigl(131 (c_1^2 + c_2^2) - 134 c_1 c_2 +  99 c_3^2 \nn\\ 
&& + ((c_1 - c_2)^2 + c_3^2) \left[28 \cos 2 \theta + \cos 4 \theta \right]\Bigr)
\eea
and
\be\label{sunpol+-}
\frac{d\sigma^{{\rm unpol.\to +-}}}{d\Omega}=\frac{\omega^6}{4(16\pi)^2}
 (c_1 + c_2)^2 \left[35 + 28 \cos 2 \theta + \cos 4 \theta  \right].
\ee
(Twice the sum of (\ref{sunpol++}) and (\ref{sunpol+-}) reproduces (\ref{sunpol}), as it should.)

\begin{figure}[b]
\centerline{\includegraphics[width=\textwidth]{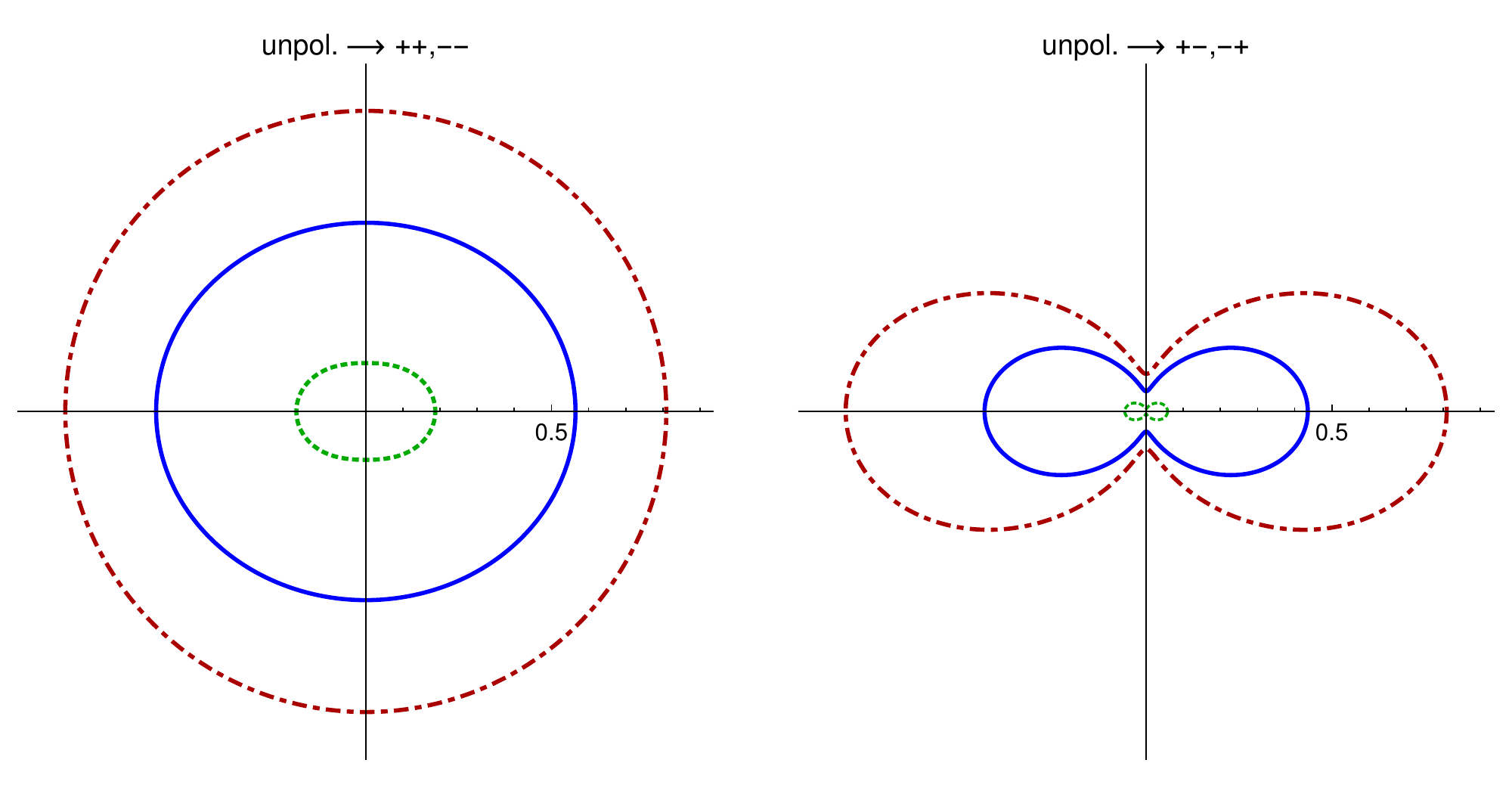}}
\caption{Leading-order differential cross section for scattering of unpolarized photons into two photons of same (left) and opposite (right)
circular polarizations, for ordinary QED (full lines), scalar QED with two charged scalar fields of mass equal to electrons (dotted lines), and supersymmetric QED (dashed-dotted lines), normalised to the maximal value of unpolarized scattering in ordinary QED.
The supersymmetric result, which has the same form as in Born-Infeld theory, turns out to be completely isotropic
in the case $unpol.\to ++,--$.
\label{fig:unpolpmpol}}
\end{figure}

The results for the three QED theories of Table \ref{tabc12} are compared in Fig.~\ref{fig:unpolpmpol}.
(In the case of scalar QED, we have doubled the matter content and considered two charged scalar fields,
because the supersymmetric case corresponds to the combination of one Dirac fermion and two scalars
as charged matter fields.)

A noteworthy feature appears in the supersymmetric/Born-Infeld case in that the differential
cross section for $\rm unpol.\to ++$ or $--$ is completely isotropic, while ordinary QED shows (a rather moderate
amount of) anisotropy. On the other hand, the result for $\rm unpol.\to +-$ or $-+$ has
a universal angular dependence.

\begin{figure}[t]
\centerline{\includegraphics[width=\textwidth]{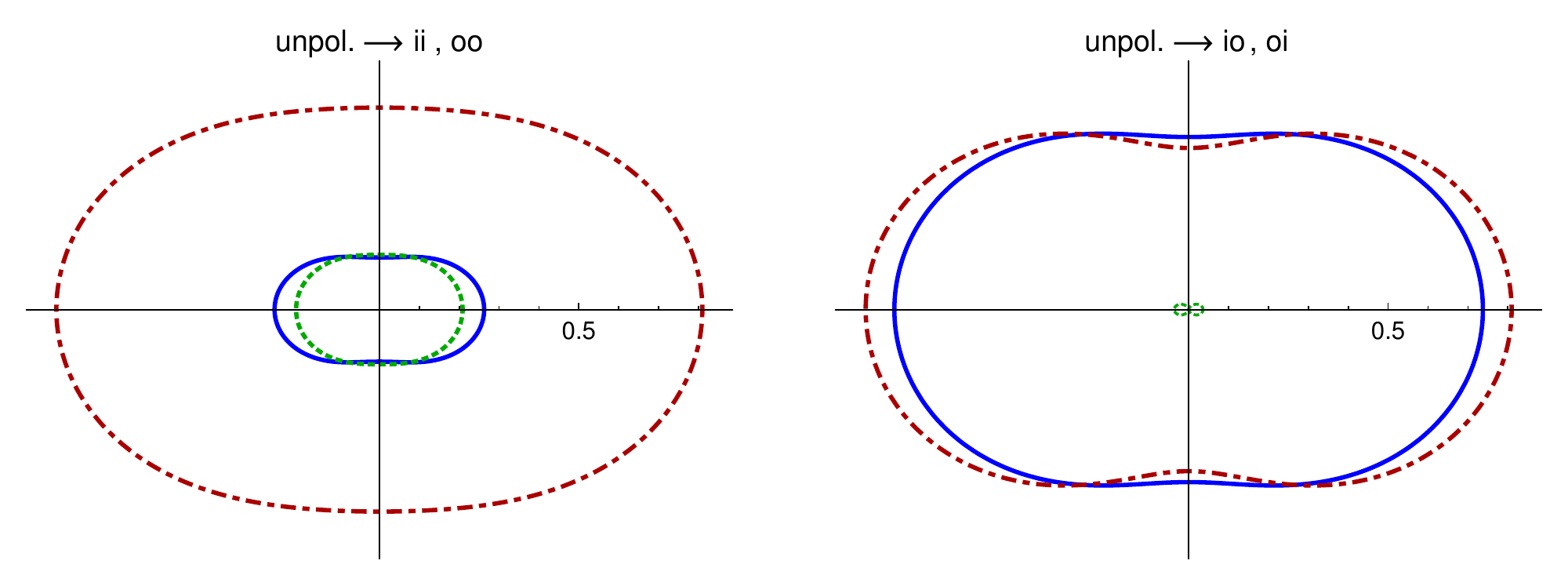}}
\caption{Same as Fig.~\ref{fig:unpolpmpol}, but for scattering into photons of same (left) and opposite (right) linear
polarizations. For same polarizations, scalar QED and ordinary QED are rather similar; for opposite polarizations
the scalar QED result is extremely suppressed, in particular at $|\theta|=\pi/2$.\label{fig:unpollinpol}}
\end{figure}

Turning next to the case of linear polarizations,
we obtain
\bea\label{sunpolii}
&&\frac{d\sigma^{{\rm unpol.\to ii}}}{d\Omega}=\frac{\omega^6}{4(16\pi)^2}
\Bigl(
262 c_1^2 - 96 c_1 c_2 + 38 c_2^2 + 99 c_3^2 \nn\\ 
&& \qquad +  4 (14 c_1^2 - 8 c_1 c_2 + 6 c_2^2 + 7 c_3^2) \cos 2 \theta + (2 (c_1^2 + c_2^2) + c_3^2) \cos 4 \theta
\Bigr)
\eea
and
\bea\label{sunpolio}
&&\frac{d\sigma^{{\rm unpol.\to io}}}{d\Omega}=\frac{\omega^6}{4(16\pi)^2}
\Bigl(
35 c_1^2 - 102 c_1 c_2 + 259 c_2^2 + 99 c_3^2 \nn\\ 
&& \qquad +  4 (7 c_1^2 - 6 c_1 c_2 + 15 c_2^2 + 7 c_3^2) \cos 2 \theta + ((c_1 - c_2)^2 + c_3^2) \cos 4 \theta
\Bigr),
\eea
which are displayed for the three versions of QED in Fig.~\ref{fig:unpollinpol}.

Now we find that for the same linear polarizations in the final state, scalar and spinor QED are rather similar in form
as well as magnitude (when both have the same number of charged degrees of freedom), but this is
completely different for opposite linear polarizations. In the latter case, the scalar QED result is extremely
suppressed, in particular for right-angle scattering $|\theta|=\pi/2$.

\subsection{Polarised initial and final states}

The total scattering cross sections of initial polarized states is given by the above expressions
through
\be
\frac{d\sigma^{\epsilon \epsilon' \to any}}{d\Omega} = 
4\frac{d\sigma^{{\rm unpol.}\to \epsilon \epsilon'}}{d\Omega}.
\ee
New features are brought about when both initial states have definite polarizations.

When all polarizations are circular, the angular dependence has universal form and only
the magnitude varies between different theories.
However, with linear polarizations, these differences become visible as different
angular patterns, occasionally involving destructive interference in certain directions.
The only exceptions are the case of all four linear polarizations being equal
and the hypothetical P and CP violating contribution involving $c_3$,
which have the same
angular dependence as the unpolarized case. However, while the 
contribution involving $c_3$ gets buried
in parity conserving contributions to
$++\to--$ and $--\to++$, with linear polarizations it would constitute the leading
low-energy contribution to scattering with an odd number of $i$ or $o$ polarizations
(if such P and CP violating vacuum polarization effects should exist).

\begin{figure}[t]
\centerline{\qquad\includegraphics[width=0.95\textwidth]{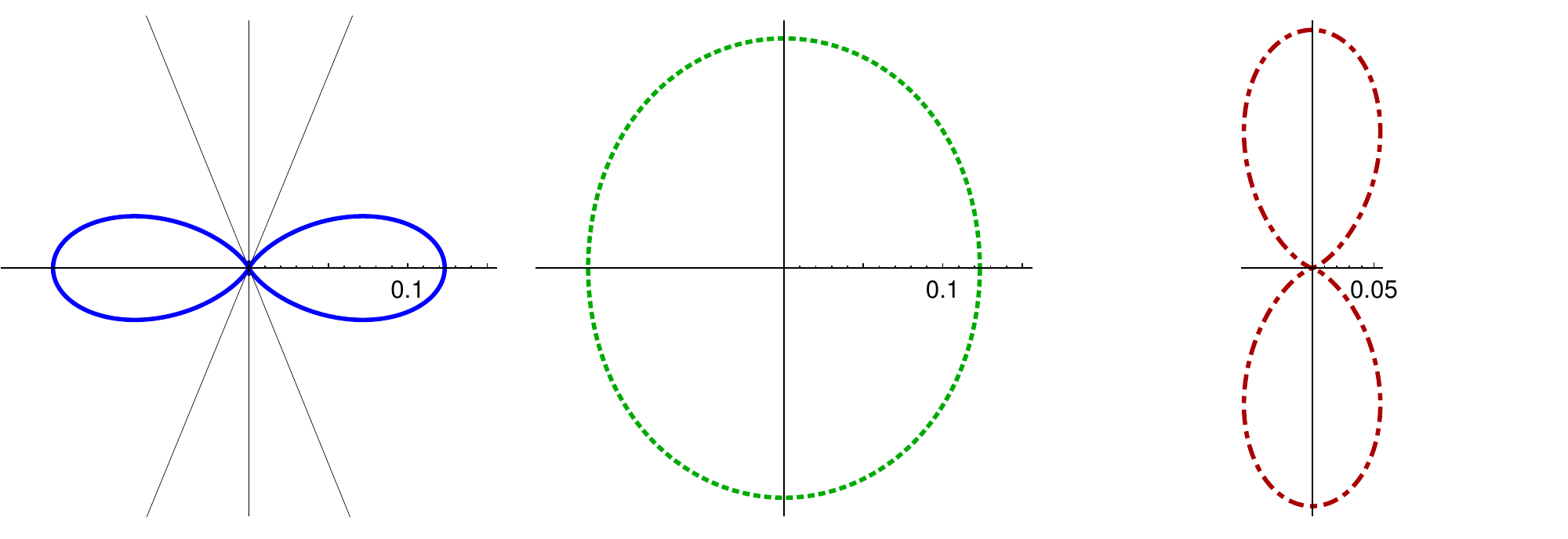}}
\centerline{\small (spinor QED)\hfil (scalar QED) \hfil (sQED/BI) }
\caption{Leading-order differential cross section for scattering of two photons with parallel linear polarizations
into two photons with polarizations orthogonal to the inital ones ($ii\to oo$ or $oo\to ii$), normalized
to the QED result for $ii\to any$ at $\theta=0$, for ordinary QED (left), scalar QED (middle), and supersymmetric QED
(right). Here the ordinary QED result has a pronounced maximum at
$\theta=0,\pi$, a secondary tiny maximum at $|\theta|=\pi/2$ (cf.\ Fig.~\ref{fig:piiooqed}), and zeros at $|\cos\theta|=1/\sqrt{7}\approx 0.378$ (denoted by thin
straight lines); the scalar QED result is maximal at $|\theta|=\pi/2$ but close to isotropic; the supersymmetric QED (Born-Infeld) result is
proportional to $\sin^4\theta$, the square of a perfect dipole.\label{fig:piioo}}
\end{figure}

\begin{figure}[h]
\centerline{\includegraphics[width=.35\textwidth]{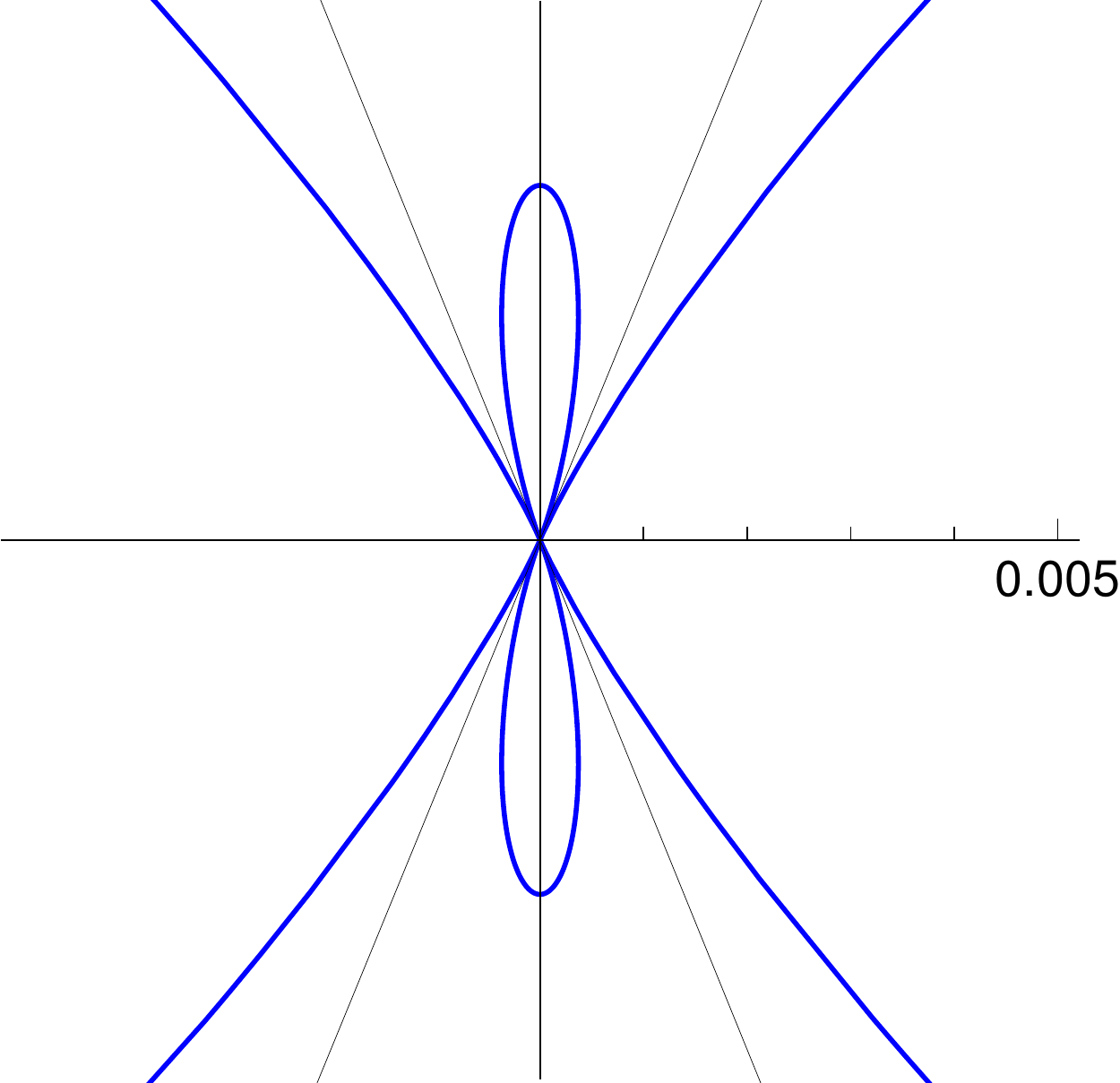}}
\caption{Magnified version of the QED result for scatterings with polarizations $ii\to oo$ or $oo\to ii$. The magnitude
of the secondary maximum at $|\theta|=\pi/2$ is smaller than the primary maximum by a factor of $1/36\approx 0.028$.\label{fig:piiooqed}}
\end{figure}

\begin{figure}[t]
\centerline{\qquad\includegraphics[width=.95\textwidth]{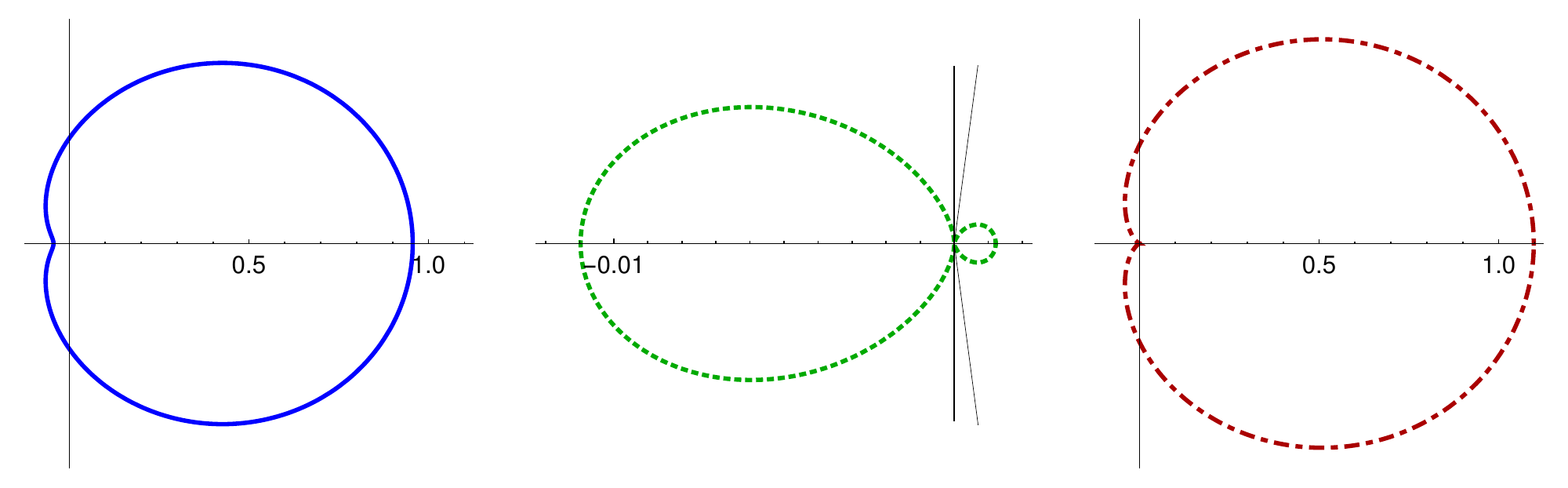}}
\centerline{\small (spinor QED)\hfil (scalar QED) \hfil (sQED/BI) }
\caption{Same as Fig.~\ref{fig:piioo}, but
for scattering of two photons with orthogonal linear polarizations
$io\to io$ or $oi\to oi$. 
The scalar QED result, which is an order of magnitude smaller than the spinor QED result, 
has zeros at $\cos\theta=(4-\sqrt{13})/3\approx 0.13$ (denoted by thin
straight lines). Its maximum at $\theta=\pi$ is exactly equal to the minimal value of the QED result there and interferes
destructively in the supersymmetric result at $\theta=\pi$.
\label{fig:pioio}}
\end{figure}

In Figures \ref{fig:piioo}--\ref{fig:pioio} we juxtapose the different patterns
for differential cross sections with processes involving linear polarizations that
are not all equal.

In Fig.~\ref{fig:piioo} the case of scattering with parallel linear polarizations is displayed,
where the final state has parallel linear polarizations orthogonal to the initial ones.
Here the three theories differ most conspicuously: ordinary spinor QED has maximal
scattering in forward and backward directions, scalar QED is roughly isotropic, and
supersymmetric QED (as well as BI electrodynamics) is maximal at right angle scattering.
However, while the latter shows a perfect squared dipole pattern, the ordinary QED result
is four lobes, with tiny lobes at right angles, made visible only in the greatly magnified
Fig.~\ref{fig:piiooqed}. At right angles
the differential cross section is a factor of $1/36\approx 0.028$
smaller than the maximal values at $\theta=0$ and $\pi$, and a
zero occurs at $|\cos\theta|=1/\sqrt{7}\approx 0.378$.

In Fig.~\ref{fig:pioio} the differential cross section for scattering with orthogonal linear polarizations
are given, $io\to io$ or $oi\to oi$. (The flipped cases $io\to oi$ or $oi\to io$ are given
by the mirror images $\theta\to\theta+\pi$.) 
Here scalar QED not only exhibits a rather different pattern than spinor and supersymmetric QED,
it is also greatly suppressed, by an order of magnitude,
similarly to the case ${\rm unpol.\to io}$ shown above.
Scalar QED has zeros at $\cos\theta=(4-\sqrt{13})/3\approx 0.13$ and a maximum for back scattering ($\theta=\pi$).
This maximum exactly equals the minimal value of the QED result at the same angle (with two charged scalars),
leading to a destructive interference at $\theta=\pi$ in the supersymmetric result.

\section{Exceptional properties of Born-Infeld electrodynamics}

The special feature of self-duality of BI electrodynamics and of supersymmetric QED 
which underlies the fact that $\mathcal M_{++++}=M_{----}^*=0$ is also
responsible for the absence of vacuum birefringence.\cite{Davila:2013wba}\footnote{In
Ref.~\citen{Hashimoto:2014dza} the absence of vacuum birefringence has been shown to hold also
for the Euler-Heisenberg Lagrangian of N=2 supersymmetric QCD at strong coupling
as derived from gauge-gravity duality.}
Vacuum birefringence means different indices of refraction for different polarizations
in the vacuum polarized by electromagnetic fields (either the field of
another wave or a constant external field), which to leading order are determined by
\be
\frac{n_1-1}{n_2-1}=\frac{c_1}{c_2}
\ee
when $c_3=0$,\footnote{See Appendix A for
the generalization to $c_3\not=0$.} 
and this equals unity for BI electrodynamics and of supersymmetric QED.

However, BI electrodynamics has further exceptional properties, some of which
go beyond supersymmetric Euler-Heisenberg Lagrangians, but which can be
understood from a string theoretic point of view.\cite{Gibbons:2000xe}

In general, a Lagrangian that is nonlinear in the two quadratic invariants $\mathcal F$
and $\mathcal G$ still has exact solutions in the form of monochromatic waves,
for which $\mathcal F=\mathcal G=0$,
independent of the size of the amplitude and thus of the degree of nonlinearity.
But it is no longer possible to superimpose different
monochromatic plane waves such that they form stable localized wave packets.
Instead, their time evolution permits singularities such as shock formation
where the limit of applicability of the effective field theory is reached.\cite{Lutzky:1959zz}

BI electrodynamics is \emph{exceptional} (in the sense of Lax) that no shocks are formed.\cite{Boillat:1970gw,Gibbons:2000xe}
A plane wave with arbitrary polarization and arbitrary profile propagating with the
speed of light is an exact solution of the field equations of BI electrodynamics\cite{Whitham}.
{\it E.g.,} the ansatz $A_y=f(t,x)$ leads to
\be
\Bigl[1+b^{-2}(f')^2\Bigr]\ddot f-2b^{-2}f' \dot f \dot f'
-\Bigl[1-b^{-2}(\dot f)^2\Bigr]f''=0,
\ee
which is solved by $f(t,x)=g(t-x)$ and $f(t,x)=g(t+x)$ with arbitrary function $g$.

In 1943, Schr\"odinger\cite{Schrodinger:1943} 
moreover found that two counter-propagating circularly polarized
monochromatic plane waves form an exact solution where the phase velocity $v$ of the two
waves in the center-of-mass system is reduced with $v^{-2}-1$ being proportional
to the energy density.
Each of the two monochromatic plane waves therefore represents a 
medium with a certain index of refraction for the other one.

A necessary condition for this to be possible can in fact be seen to hold in
the above results for polarized differential cross sections.
In BI electrodynamics we have $\mathcal M_{++++}=0$, which actually means
that there is no scattering of two right-handed photons into left-handed ones,
$\sigma(++\to --)=0$;
BI electrodynamics preserves helicity\cite{Rosly:2002jt}.
The so-called\cite{Bern:1994zx}\footnote{Amplitudes with two helicities of one type and all the others
of the other type are called maximally helicity-violating (MHV) because the even
more helicity-violating ones with all helicity indices equal, or only one unequal,
vanish in a \emph{supersymmetric} gauge theory, and have no cuts for general gauge theories.}
MHV amplitude $\mathcal M_{++--}\not=0$ corresponds (here slightly confusingly) to $++\to++$,
and this is isotropic. The situation is therefore
similar to electrodynamics in an isotropic medium with a polarization-independent
refractive index.
The homogeneous isotropic scattering superimposed on
the individual plane wave traveling originally with the speed of light reduces its phase velocity uniformly. In the case of light-by-light scattering a counter-propagating wave
of same helicity effectively provides such an isotropic medium.



\section{Conclusion}

In this paper we have studied the general form of polarization-dependent
differential cross sections of light-by-light scattering at low energies,
even including a parity violating contribution.
While the angular dependence of all amplitudes with given helicity states
has a universal form such that only their magnitude varies between different
effective field theories (with a zero for $\mathcal M_{++++}=\mathcal M_{----}^*$
in the case of supersymmetric QED and Born-Infeld electrodynamics),
linear polarizations lead to interesting patterns for different theories.
Moreover, a parity violating contribution, which in the case of circular
polarizations is buried in the parity conserving contributions
of (non-supersymmetric) QED
to polarizations $++\to--$ and $--\to++$, appears as a leading contribution
in scattering with an odd number of parallel linear polarizations.

\section*{Acknowledgments}

A.R.\ would like to thank Holger Gies, Ayan Mukhopadhyay, and Peter van Nieuwenhuizen 
for useful discussions, and Arkady Tseytlin for pointing out an incorrect
statement in footnote c in the previous (published) version of this paper.

\appendix
\section{Generalized formulae for vacuum birefringence}

The continuous scattering of light in a background provided by extended and strong coherent
electromagnetic fields can be described by refractive indices, which in general
depend on polarization. This gives rise to the phenomenon of vacuum birefringence\cite{Klein:1964,Baier:1967zzc,Brezin:1971nd},
for which recently direct evidence has been claimed in the optical
polarimetry observation of the isolated neutron star RX\,J1856.5$-$3754.\cite{Mignani:2016fwz}

In this Appendix we extend the analysis of Ref.~\citen{Heinzl:2006xc} to
the general form of the Lagrangian for nonlinear electrodynamics (\ref{L4lowen}), including
the P and CP violating term proportional to $c_3$.

Linearizing the field equations following from (\ref{L4lowen}) in the presence of
electromagnetic background fields, $F_{\mu\nu}\to F_{\mu\nu}+(\partial_\mu A_\nu-\partial_\nu A_\mu)$, 
one obtains the following fluctuation equation for the potential $A^\mu(k)$
in momentum space,
\be\label{linfeq}
(1-2 c_1\mathcal F-c_3 \mathcal G)(g_{\mu\nu}k^2-k_\mu k_\nu)A^\nu
=\left[2 c_1 b_\mu b_\nu+2 c_2 \tilde b_\mu \tilde b_\nu
+c_3(b_\mu \tilde b_\nu+\tilde b_\mu b_\nu)\right]A^\nu
\ee
with
\be
b_\mu=F_{\mu\nu}k^\nu, \quad \tilde b_\mu=\tilde F_{\mu\nu}k^\nu,
\ee
satisfying $b\cdot k=0=\tilde b\cdot k$.
Nontrivial background fields lead to nonzero $k^2\sim {\rm max\,}(c_1,c_2,c_3)$ ($\propto \alpha^2$ in QED).
To first order\footnote{A more general analysis of birefringence effects
in nonlinear electrodynamics (likewise permitting parity violating terms) can be
found in Ref.~\citen{Obukhov:2002xa}.}
in the $c$'s, one can drop the terms involving $\mathcal F$ and $\mathcal G$
on the left-hand side of (\ref{linfeq}). Furthermore, since $k^\mu$ is approximately light-like, we
also have $b\cdot \tilde b=0$ and $b^2=\tilde b^2=-\omega^2 Q^2$ with a nonnegative quantity\cite{Heinzl:2006xc}
$Q^2$. The latter equals $(\ve_k \times \vB)^2$ in a constant magnetic background field; in
a counter-propagating plane wave one has $Q^2=4I$, with $I$ denoting the background energy density.

With $k^\mu=\omega(1,n\ve_k)$, 
(\ref{linfeq}) has eigenvector solutions $A^\nu\propto\beta b^\nu+\tilde\beta \tilde b^\nu$
with two possible values for the refractive index $n$,
\be
n_{1,2}^2-1=\left[c_1+c_2\pm \sqrt{(c_1-c_2)^2+c_3^2}\right] Q^2.
\ee
With $c_3=0$ this reduces to $n_{1,2}\approx 1+c_{1,2}Q^2$ and $A^\nu_1\propto b^\nu, A^\nu_2\propto \tilde b^\nu$.
When $c_3\not=0$, $b^\nu$ and $\tilde b^\nu$ get mixed with an angle $\delta$ given by
\be
\tan\delta=\frac{\beta_2}{\tilde \beta_2}=-\frac{\tilde\beta_1}{\beta_1}=
\frac{c_3}{c_1-c_2+\sqrt{(c_1-c_2)^2+c_3^2}}\,.
\ee

%
%
%
%
%
%
%
%

\bibliographystyle{ws-ijmpa}
\bibliography{RT}

\begin{thebibliography}{10}
\expandafter\ifx\csname urlstyle\endcsname\relax
  \providecommand{\doi}[1]{doi:\discretionary{}{}{}#1}\else
  \providecommand{\doi}{doi:\discretionary{}{}{}\begingroup
  \urlstyle{rm}\Url}\fi

\bibitem{Euler:1935zz}
H.~Euler and B.~Kockel, {\em Naturwiss.} {\bf 23},   246  (1935).

\bibitem{Euler:1936}
H.~Euler, {\em Ann. d. Phys.} {\bf 26}, 398  (1936).

\bibitem{Akiezer1936}
A.~Akhieser, L.~Landau and I.~Pomeranchook, {\em Nature} {\bf 138},   206
  (1936).

\bibitem{Akiezer1937}
A.~Achieser, {\em Physik Z. Sowjetunion} {\bf 11},   263  (1937).

\bibitem{Karplus:1950zz}
R.~Karplus and M.~Neuman, {\em Phys. Rev.} {\bf 83}, 776  (1951).

\bibitem{Heisenberg:1935qt}
W.~Heisenberg and H.~Euler, {\em Z. Phys.} {\bf 98},   714  (1936).

\bibitem{Weisskopf:1936}
V.~Weisskopf, {\em Kong. Dan. Vid. Selsk. Mat.-fys. Medd.} {\bf XIV/6}, 1
  (1936).

\bibitem{Dittrich:2000zu}
W.~Dittrich and H.~Gies, {\em Springer Tracts Mod. Phys.} {\bf 166}, 1  (2000).

\bibitem{Dunne:2004nc}
G.~V. Dunne, Heisenberg-{E}uler effective lagrangians: Basics and extensions,
  in {\em From Fields to Strings: Circumnavigating Theoretical Physics - Ian
  Kogan Memorial Collection\/},  eds. M.~Shifman, A.~Vainshtein and J.~Wheater
  (World Scientific, Singapore, 2004) pp. 445--522, 
\newblock \href{http://arxiv.org/abs/hep-th/0406216}{{\ttfamily
  arXiv:hep-th/0406216}}.

\bibitem{Born:1934gh}
M.~Born and L.~Infeld, {\em Proc. Roy. Soc. Lond.} {\bf A144}, 425  (1934).

\bibitem{Schrodinger:1942}
E.~Schr{\"o}dinger, {\em Proc. Roy. Irish Acad. (A)} {\bf 47}, 77  (1942).

\bibitem{Schrodinger:1943}
E.~Schr{\"o}dinger, {\em Proc. Roy. Irish Acad. (A)} {\bf 49}, 59  (1943).

\bibitem{Fradkin:1985qd}
E.~S. Fradkin and A.~A. Tseytlin, {\em Phys. Lett.} {\bf B163}, 123  (1985).

\bibitem{Cecotti:1986gb}
S.~Cecotti and S.~Ferrara, {\em Phys. Lett.} {\bf B187}, 335  (1987).

\bibitem{Tseytlin:1986ti}
A.~A. Tseytlin, {\em Nucl. Phys.} {\bf B276},   391  (1986), [Erratum: Nucl.
  Phys.B291, 876 (1987)].

\bibitem{Metsaev:1987qp}
R.~R. Metsaev, M.~Rakhmanov and A.~A. Tseytlin, {\em Phys. Lett.} {\bf B193},
  207  (1987).

\bibitem{Tseytlin:1999dj}
A.~A. Tseytlin, { {Born-Infeld action, supersymmetry and string theory}}, in
  {\em The Many Faces of the Superworld - Yuri Golfand Memorial Volume\/},  ed.
  M.~A. Shifman (World Scientific, Singapore, 1999), pp. 417--452,
\newblock \href{http://arxiv.org/abs/hep-th/9908105}{{\ttfamily
  arXiv:hep-th/9908105}}.

\bibitem{Gibbons:1997xz}
G.~W. Gibbons, {\em Nucl. Phys.} {\bf B514}, 603  (1998),
  \href{http://arxiv.org/abs/hep-th/9709027}{{\ttfamily arXiv:hep-th/9709027}}.

\bibitem{Gibbons:2000xe}
G.~W. Gibbons and C.~A.~R. Herdeiro, {\em Phys. Rev.} {\bf D63},   064006
  (2001), \href{http://arxiv.org/abs/hep-th/0008052}{{\ttfamily
  arXiv:hep-th/0008052}}.

\bibitem{Marklund:2008gj}
M.~Marklund and J.~Lundin, {\em Eur. Phys. J.} {\bf D55}, 319  (2009),
  \href{http://arxiv.org/abs/0812.3087}{{\ttfamily arXiv:0812.3087 [hep-th]}}.

\bibitem{Heinzl:2006xc}
T.~Heinzl, B.~Liesfeld, K.-U. Amthor, H.~Schwoerer, R.~Sauerbrey and A.~Wipf,
  {\em Opt. Commun.} {\bf 267}, 318  (2006),
  \href{http://arxiv.org/abs/hep-ph/0601076}{{\ttfamily arXiv:hep-ph/0601076}}.

\bibitem{Karbstein:2015xra}
F.~Karbstein, H.~Gies, M.~Reuter and M.~Zepf, {\em Phys. Rev.} {\bf D92},
  071301  (2015), \href{http://arxiv.org/abs/1507.01084}{{\ttfamily
  arXiv:1507.01084 [hep-ph]}}.

\bibitem{Klein:1964}
J.~J. Klein and B.~P. Nigam, {\em Phys. Rev.} {\bf 135}, B1279  (1964).

\bibitem{Baier:1967zzc}
R.~Baier and P.~Breitenlohner, {\em Nuovo Cim.} {\bf 47}, 117  (1967).

\bibitem{Brezin:1971nd}
E.~Brezin and C.~Itzykson, {\em Phys. Rev.} {\bf D3}, 618  (1971).

\bibitem{Fouche:2016qqj}
M.~Fouch\'e, R.~Battesti and C.~Rizzo, {\em Phys. Rev.} {\bf D93},   093020
  (2016), \href{http://arxiv.org/abs/1605.04102}{{\ttfamily arXiv:1605.04102
  [physics.optics]}}.

\bibitem{Duff:1979bk}
M.~J. Duff and C.~J. Isham, {\em Phys. Lett.} {\bf B86}, 157  (1979).

\bibitem{Grisaru:1976vm}
M.~T. Grisaru, H.~N. Pendleton and P.~van Nieuwenhuizen, {\em Phys. Rev.} {\bf
  D15},   996  (1977).

\bibitem{Schrodinger:1935oqa}
E.~Schr{\"o}dinger, {\em Proc. Roy. Soc. Lond.} {\bf A150}, 465  (1935).

\bibitem{BialynickiBirula:1984tx}
I.~Bia{\l}ynicki-Birula, { Nonlinear electrodynamics: Variations on a theme by
  {B}orn and {I}nfeld}, in {\em Quantum theory of particles and fields -
  Birthday volume dedicated to Jan {\L}opusza{\'n}ski\/},  eds. B.~Jancewicz
  and J.~Lukierski (World Scientific, Singapore, 1983), pp. 31--48.

\bibitem{Gibbons:1995cv}
G.~W. Gibbons and D.~A. Rasheed, {\em Nucl. Phys.} {\bf B454}, 185  (1995),
  \href{http://arxiv.org/abs/hep-th/9506035}{{\ttfamily arXiv:hep-th/9506035}}.

\bibitem{Gaillard:1997zr}
M.~K. Gaillard and B.~Zumino, { Selfduality in nonlinear electromagnetism}, in
  {\em Supersymmetry and quantum field theory - Proceedings of the D. Volkov Memorial
  Seminar, Kharkov, Ukraine, January 5-7, 1997\/}, Lect. Notes Phys. Vol.~509
  (1998), pp. 121--129,
\newblock \href{http://arxiv.org/abs/hep-th/9705226}{{\ttfamily
  arXiv:hep-th/9705226}}.

\bibitem{Kuzenko:2007cg}
S.~M. Kuzenko and S.~J. Tyler, {\em JHEP} {\bf 05},   081  (2007),
  \href{http://arxiv.org/abs/hep-th/0703269}{{\ttfamily arXiv:hep-th/0703269}}.

\bibitem{Davila:2013wba}
J.~M. D{\'a}vila, C.~Schubert and M.~A. Trejo, {\em Int. J. Mod. Phys. A} {\bf
  29},   1450174  (2014), \href{http://arxiv.org/abs/1310.8410}{{\ttfamily
  arXiv:1310.8410 [hep-ph]}}.

\bibitem{IZ}
C.~Itzykson and J.-B. Zuber, {\em Quantum field theory} (McGraw-Hill,
  Singapore, 1985).

\bibitem{Liang:2011sj}
Y.~Liang and A.~Czarnecki, {\em Can. J. Phys.} {\bf 90}, 11  (2012),
  \href{http://arxiv.org/abs/1111.6126}{{\ttfamily arXiv:1111.6126 [hep-ph]}}.

\bibitem{LL4}
E.~Lifshitz, V.~Berestetskii and L.~Pitaevskii, {\em Quantum electrodynamics}
  (Butterworth-Heinemann, Oxford, 1982).

\bibitem{Hashimoto:2014dza}
K.~Hashimoto, T.~Oka and A.~Sonoda, {\em JHEP} {\bf 06},   085  (2014),
  \href{http://arxiv.org/abs/1403.6336}{{\ttfamily arXiv:1403.6336 [hep-th]}}.

\bibitem{Lutzky:1959zz}
M.~Lutzky and J.~S. Toll, {\em Phys. Rev.} {\bf 113}, 1649  (1959).

\bibitem{Boillat:1970gw}
G.~Boillat, {\em J. Math. Phys.} {\bf 11}, 941  (1970).

\bibitem{Whitham}
G.~B. Whitham, {\em Linear and nonlinear waves} (Wiley, New York, 1974).

\bibitem{Rosly:2002jt}
A.~A. Rosly and K.~G. Selivanov, { {Helicity conservation in Born-Infeld
  theory}}, in {\em {Quarks. Proceedings, 12th International Seminar on High
  Energy Physics, Quarks'2002, Novgorod, Russia, June 1-7, 2002}\/},  (2002),
\newblock \href{http://arxiv.org/abs/hep-th/0204229}{{\ttfamily
  arXiv:hep-th/0204229}}.

\bibitem{Bern:1994zx}
Z.~Bern, L.~J. Dixon, D.~C. Dunbar and D.~A. Kosower, {\em Nucl. Phys.} {\bf
  B425}, 217  (1994), \href{http://arxiv.org/abs/hep-ph/9403226}{{\ttfamily
  arXiv:hep-ph/9403226}}.

\bibitem{Mignani:2016fwz}
R.~P. Mignani, V.~Testa, D.~G. Caniulef, R.~Taverna, R.~Turolla, S.~Zane and
  K.~Wu, {\em Mon. Not. R. Astron. Soc.} {\bf 465}, 492  (2016),
  \href{http://arxiv.org/abs/1610.08323}{{\ttfamily arXiv:1610.08323
  [astro-ph.HE]}}.

\bibitem{Obukhov:2002xa}
Y.~N. Obukhov and G.~F. Rubilar, {\em Phys. Rev.} {\bf D66},   024042  (2002),
  \href{http://arxiv.org/abs/gr-qc/0204028}{{\ttfamily arXiv:gr-qc/0204028}}.

\end{thebibliography}
\end{document}